\documentstyle[epsfig,rotating,12pt]{elsart}

\def\rpv{$R_p \hspace{-1em}/\;\:$ }
\def\eslash{$E \hspace{-0.7em}/\;\:$ }
\def\gsim{\raise0.3ex\hbox{$\;>$\kern-0.75em\raise-1.1ex\hbox{$\sim\;$}}}
\def\lsim{\raise0.3ex\hbox{$\;<$\kern-0.75em\raise-1.1ex\hbox{$\sim\;$}}}

\begin{document}

\begin{flushright}hep-ph/0503059 \\ IFIC/05-18 \\ ZU-TH 04/05 \end{flushright}

\begin{frontmatter}

\vspace*{0cm} \title{Collider signals of gravitino dark matter in 
bilinearly broken R-parity}

\author{M.~Hirsch, $^{\$}$ \thanksref{mahirsch}}
\author{W.~Porod $^{\$,*}$ \thanksref{porod}} and 
\author{D.~Restrepo $^{\$,**}$\thanksref{restrepo}}

\address{$^{\$}$ Instituto de
    F\'{\i}sica Corpuscular / C.S.I.C. - Universitat de Val\`encia,
    \\ Edificio Institutos de Paterna, Apartado de Correos 22085 \\ 
    46071 Val\`encia, Spain}

\address{$^*$ Institut f\"ur Theoretische Physik,
      Universit\"at Z\"urich, Switzerland}

\address{$^{**}$ Insituto de F\'{i}sica, Universidad de Antioquia, \\
A.A 1226, Medellin, Colombia}

\thanks[mahirsch]{mahirsch@ific.uv.es}
\thanks[porod]{porod@ific.uv.es}
\thanks[restrepo]{restrepo@ific.uv.es}



\begin{abstract}

In models with gauge mediated supersymmetry breaking the gravitino is 
the lightest supersymmetric particle. If R-parity is violated the 
gravitino decays, but with a half-live far exceeding the age of the 
universe and thus is, in principle, a candidate for the dark matter. 
We consider the decays of the next-to-lightest supersymmetric particle, 
assumed to be the neutralino. We show that in models where the 
breaking of R-parity is bilinear, the condition that R-parity violation 
explains correctly the measured neutrino masses fixes the branching ratio 
of the decay ${\tilde \chi}^0_1 \rightarrow {\tilde G}\gamma$ in the range 
$10^{-3}-10^{-2}$, if the gravitino mass is in the range required to 
solve the dark matter problem, i.e. of the order (few) $100$ eV. This 
scenario is therefore directly testable at the next generation of colliders.

\end{abstract}  

\vskip5mm
\hrule

\smallskip
{\it \small PACs: 14.60.Pq, 12.60.Jv, 95.35.+d}

\maketitle

\end{frontmatter}

\section{Introduction
        \label{Introduction}}

Models of gauge mediated supersymmetry breaking (GMSB) generically predict 
that the lightest supersymmetric particle (LSP) is the gravitino 
\cite{Giudice:1998bp}. Such a light gravitino, in principle, is a candidate 
for the non-baryonic dark matter of the universe \cite{Pagels:1981ke}. 
The smallness of the gravitino couplings, however, make such a scenario 
extremely difficult to test. Direct detection of gravitino dark matter 
in scattering experiments or indirectly via decays/annihilation to gamma 
rays is hopeless \cite{Bertone:2004pz} and consequently gravitino dark 
matter has received rather scarce attention. 

The purpose of the present letter is to show that in models with bilinear 
breaking of R-parity the branching ratio of the decay of the neutralino 
into a gravitino and a photon is fixed by data on neutrino masses up 
to a factor of $m_{3/2}^{-2}$, if the neutralino is the next-to lightest 
supersymmetric particle (NLSP). A measurement of this branching ratio 
thus implies a ``measurement'' of the gravitino mass, $m_{3/2}$. 
Approximate knowledge of $m_{3/2}$ in turn can be used to constrain the 
conjecture that the gravitino is the (major component of the) dark matter 
in the universe.

On the one hand, supersymmetric models with R-parity violation can 
explain \cite{hirsch:2000ef,hempfling:1996wj,drees:1998id,davidson:2000ne} 
current data on neutrino masses and mixings \cite{Maltoni:2004ei} without 
invoking any GUT-scale physics. On the other hand, in supersymmetric models 
with R-parity violation the LSP decays. For all superpartners of standard 
model particles these decays proceed at rates that even the most tiny amount 
of R-parity violation rules out MSSM particles as dark matter. A light 
gravitino, however, couples so weakly to standard model particles that its 
half-live far exceeds the age of the universe even for R-parity violating 
couplings as large as ${\cal O}(1)$ \cite{Borgani:1996ag,Takayama:2000uz}, 
see also section 2. Thus, contrary to popular believe, supersymmetry holds 
the promise to solve the dark matter problem even if R-parity is violated.

If gravitinos were in thermal equilibrium in the early universe (and 
assuming that there is no non-standard physics between gravitino decoupling 
and the time of nucleosynthesis, see below), the contribution of gravitinos 
to the matter content of the universe can be estimated to be 
\cite{Pagels:1981ke} 
\begin{equation}
\Omega_{3/2}h^2 \simeq 0.11 \Big(\frac{m_{3/2}}{\rm 100 \hskip2mm eV}\Big)
\Big(\frac{100}{g_{*}}\Big).
\label{OmGrav}
\end{equation}
Here, $\Omega_{3/2}$ is the density of gravitinos in units of the critical 
density, $h$ is the Hubble parameter in units of 100 km s$^{-1}$ Mpc$^{-1}$ 
and $g_{*}$ is the effective number of degrees of freedom at the time of 
gravitino decoupling. Depending on the so-far unknown supersymmetric 
particle spectrum one expects $g_{*} \simeq 90-140$ \cite{Pierpaoli:1997im}. 
Current data give the matter density of the universe \cite{Eidelman:2004wy} 
as $\Omega_{M} h^2 \simeq 0.134 \pm 0.006$, from which 
$\Omega_{B} h^2 \simeq 0.023 \pm 0.001$ is in the form of baryons. 

Particle dark matter (DM) is usually classified according to its 
free-streaming length \cite{Primack:1997av} as either ``hot'', ``warm'' 
or ``cold'' DM. There is a general consensus that hot DM is ruled 
out \cite{Bertone:2004pz,Eidelman:2004wy,Primack:1997av}. Cold DM is 
usually considered the best choice \cite{Bertone:2004pz,Primack:1997av} 
to fit large-scale structure data. However, on galactic and sub-galactic 
scales pure cold DM seems to produce too much power, see for example 
\cite{Moore:1999gc,Bode:2000gq} and references therein. 
\footnote{For a different point of view, see for example 
\cite{Swaters:2002rx}.} To resolve the deficits of cold DM, some groups 
considered warm DM variants, claiming that WDM  does actually provide a 
better fit \cite{Bode:2000gq,Avila-Reese:2000hg}. However, constraints on 
the free streaming length (and thus the mass) of WDM particles can be 
derived from data of the Lyman-$\alpha$ forest \cite{Narayanan:2000tp}, 
and a lower limit of $m_{WDM} \ge 0.55$ keV for {\em thermal} relics 
is quoted in \cite{Viel:2005qj}.

The lower limit on the mass of WDM particles given in \cite{Viel:2005qj} 
seems to be in conflict with the conjecture that gravitinos are the 
dark matter, see Eq.(\ref{OmGrav}). However, in the derivation of the 
gravitino density \cite{Pagels:1981ke} it is assumed that the universe 
has a ``standard'' thermal history. Producing additional entropy after 
the time of gravitino decoupling would dilute the density of gravitinos, 
\footnote{The other logical possibility, i.e. to raise $g_{*}$ to values 
of the order of $g_{*} \simeq (600-700)$ by the introduction of a sea 
of new particles, does not seem very economical.} compared to the estimate 
Eq.(\ref{OmGrav}), two variations of this idea are discussed in 
\cite{Baltz:2001rq,Fujii:2002fv}. Both \cite{Baltz:2001rq} and 
\cite{Fujii:2002fv} assume that entropy is produced by the ``late'' decay 
of messenger particles. \footnote{It has been speculated that messenger 
particles themselves might provide the dark matter, see for example 
\cite{Dimopoulos:1996gy,Han:1997wn}. However, messengers tend to overclose 
the universe unless the mass of the lightest messenger is rather low, of 
the order of $m_{LMP} \simeq {\cal O}(1)$ TeV.} Baltz and Murayama 
\cite{Baltz:2001rq} argue that the lightest messenger particle might decay 
through an intermediate heavy particle with mass $m_X \simeq 10^{12}$ GeV, 
which leads to a messenger decay width sufficiently small to dilute the 
gravitino density by a factor of ($5-8$). Fujii and Yanagida, on the 
other hand, claim that adding a constant messenger number 
violating term to the superpotential, messenger widths of the ``correct'' 
order of magnitude are naturally obtained \cite{Fujii:2002fv}. 

Given this discussion, we think it is fair to say that 
gravitinos with a mass in the range of ${\cal O}(0.1)$-${\cal O}(1)$ 
keV are interesting dark matter candidates. Constraining the gravitino 
mass to be much smaller than given by Eq.(\ref{OmGrav}) would rule out 
gravitinos as DM. 

NLSP decays in R-parity violating variants of GMSB have been 
considered previously \cite{Carena:1997wy}. For the case of bilinear 
R-parity breaking the authors of \cite{Carena:1997wy} point out that a 
bound on $|\vec\Lambda|/\sqrt{{\rm det}{\cal M}_{{\tilde \chi}^0}}$ of 
(very roughly) 
the order ${\cal O}(10^{-6})$ for $\sqrt{F}=10^6$ GeV can be obtained from 
the requirement $\Gamma({\tilde \chi}^0_1\rightarrow {\tilde G}\gamma) \ge 
\sum \Gamma({\tilde \chi}^0_1\rightarrow$ \rpv $)$. Fits \cite{hirsch:2000ef} 
to current neutrino data \cite{Maltoni:2004ei} require similar, although 
somewhat larger, values for bilinear R-parity violating parameters, see 
next section. In our numerical calculation we thus find 
$\Gamma({\tilde \chi}^0_1\rightarrow {\tilde G}\gamma)/ 
\sum \Gamma({\tilde \chi}^0_1\rightarrow$ \rpv $) < 1$, unless the 
gravitino mass is much smaller than indicated in eq. (\ref{OmGrav}). 

The remainder of this paper is organized as follows. In the next section we 
present some approximate formulas for the decay of the neutralino NLSP. 
These estimates serve to understand the results of the numerical analysis, 
presented next. We then close with a short summary. 

\section{Semi-analytical estimates}

In this section we give some (semi-) analytical formulas for the decay 
of a neutralino NLSP. This will facilitate the understanding of our 
numerical results presented below. In GMSB with R-parity violation, a 
neutralino NLSP can either decay into a gravitino and a photon or via 
R-parity violating interactions directly to standard model particles.

We take into account only bilinear R-parity violating (BRpV) terms, 
namely,
\begin{eqnarray}\nonumber 
W = W_{MSSM} + \epsilon_i {\widehat L}_i {\widehat H}_u, \\
V_{\rm soft} = V^{MSSM}_{\rm soft} + B_i \epsilon_i {\tilde L}_i H_u.
\label{DefModel}
\end{eqnarray}
Eq. (\ref{DefModel}) can be considered as a minimal model of R-parity 
violation. The new terms in $V_{\rm soft}$ induce vacuum expectation 
values for the scalar neutrinos $v_i$. One can either treat $B_i$ or 
$v_i$ as free parameters of the model, since they are connected by 
the tadpole equations. 

In models with bilinear breaking of R-parity, one neutrino mass is 
generated at tree-level, while the other neutrino masses are due to 
1-loop corrections \cite{hirsch:2000ef}. The decay width of the NLSP 
to standard model particles is related to the neutrino masses, thus 
we first discuss some approximate formulas for the calculation of neutrino 
masses in BRpV models.

The tree-level contribution of BRpV to neutrino masses is given as
\begin{equation}
m_{\nu}^{\rm tree} = \frac{m_{\tilde \gamma}}
   {4 {\rm det}{\cal M}_{{\tilde \chi}^0}}|{\vec \Lambda}|^2
\label{mTree}
\end{equation}
Here, $m_{\tilde \gamma}$ is the ``photino mass'' $m_{\tilde \gamma} =
g^2 M_1 + g'^2 M_2$, ${\rm det}{\cal M}_{{\tilde \chi}^0}$ is the determinant 
of the ($4 \times 4$) MSSM neutralino mass matrix and ${\vec \Lambda}$ is the 
so-called alignment vector, $\Lambda_i = \epsilon_i v_d + v_i \mu$. 

The dominant 1-loop corrections to the neutrino mass matrix are usually 
due to bottom/sbottom and tau/stau loops and are, very roughly, of 
order \cite{hirsch:2000ef}
\begin{equation}
\label{Simplest}
m^{\rm 1lp}_{\nu} 
\simeq \frac{1}{16 \pi^2} \Big( 3 h_b^2 \sin(2\theta_{\tilde b}) 
m_b \Delta B_0^{\tilde b_2\tilde b_1} + h_{\tau}^2 \sin(2\theta_{\tilde \tau}) 
m_{\tau} \Delta B_0^{\tilde \tau_2\tilde \tau_1} \Big)
\frac{({\tilde \epsilon}_1^2 + {\tilde \epsilon}_2^2)}{\mu^2}.
\end{equation}
Here, $h_b$ ($h_{\tau}$) are the bottom ($\tau$) Yukawa coupling, 
$\theta_{\tilde b}$ ($\theta_{\tilde \tau}$) is the mixing angle 
in the sbottom (stau) sector, $\Delta B_0^{ab}$ is the difference of 
two Passarino-Veltman $B_0$-functions, essentially 
$\Delta B_0^{ab} \simeq {\rm ln}(m_a^2/m_b^2)$ for sfermion masses much 
larger than the corresponding fermion masses. And, finally 
${\tilde \epsilon}$ are the superpotential parameters $\vec\epsilon$ 
rotated to the basis where the tree-level neutrino mass matrix is 
diagonal. 

Eqs (\ref{mTree}) and (\ref{Simplest}) produce a hierarchical neutrino 
spectrum. We will assume that the tree-level contribution is larger 
than the 1-loop correction. Thus, we identify $m_{\nu}^{\rm tree} 
\simeq \sqrt{\Delta m^2_{Atm}} \simeq 0.04-0.06$ eV and 
$m^{\rm 1lp}_{\nu} \simeq \sqrt{\Delta m^2_{\odot}} \simeq 0.009$ eV.
For any given choice of R-parity conserving SUSY parameters then 
$|\vec\Lambda|/\sqrt{{\rm det}{\cal M}_{{\tilde \chi}^0}}$ and 
$|\vec\epsilon|/\mu$ are approximately fixed by neutrino masses. 
Typical values are \cite{hirsch:2000ef}:  
$|\vec\Lambda|/\sqrt{{\rm det}{\cal M}_{{\tilde \chi}^0}} \sim$ (few) 
10$^{-6}$ and $|\vec\epsilon|/\mu  \sim$ (few) 10$^{-4}$. 

The neutralino will decay to three SM fermion final states or, if 
kinematically allowed, into gauge bosons and leptons, $W^{\pm}l^{\mp}$ 
and $Z^0\nu$. \footnote{The decay to a light Higgs plus neutrinos, $h^0\nu$, 
is also possible. However, for generic GMSB parameter choices it is 
less important.} To estimate the most important decay widths, we will 
make use of the approximate neutralino couplings in first order expansion 
in small \rpv parameters as given in \cite{Porod:2000hv}.

Consider first the decay to gauge bosons. In GMSB scenarios 
the lightest neutralino is usually bino dominated \cite{Giudice:1998bp}. 
Binos couple to gauge bosons proportional to $\Lambda_i$. With couplings 
from \cite{Porod:2000hv} and Eq.(\ref{mTree}) we estimate
\begin{eqnarray} 
\Gamma({\tilde \chi}^0_1 \rightarrow \sum_i W^{\pm}l^{\mp}) 
\sim & &
\frac{g^2 g'^2 M_2 m_{{\tilde \chi}^0_1} }
{(16\pi M_1 m_{\gamma})} 
f(m_{W}^2/m_{{\tilde \chi}^0_1}^2)m_{\nu}^{\rm Tree}.
\label{GamRPV}
\end{eqnarray}
Here, $f(x)$ is a phase space factor, given by 
$f(x) = \frac{1}{2 x} - \frac{3 x}{2} +x^2$. A similar expression 
holds for $\Gamma({\tilde \chi}^0_1 \rightarrow \sum_i Z^0 \nu_i)$ with an 
additional prefactor of $1/(4 c_W^2)$. Assuming $M_1 \simeq M_2/2$ 
and $|\mu|/M_1 \simeq 4$, as is typical for GMSB models, results very 
roughly in $\Gamma \simeq 2 \times 10^{-4} \frac{m_{\nu}}{\rm 0.05 eV} 
f(m_{W}^2/m_{{\tilde \chi}^0_1}^2)$ eV.

Neutralino decays to three fermions can also be mediated by scalar 
quark and scalar lepton exchange. With approximate scalar lepton couplings 
from \cite{Hirsch:2002ys} we estimate that the decay 
$ \Gamma({\tilde \chi}^0_1 \rightarrow \nu \tau^{\pm}l^{\mp})$ is very 
roughly of order
\begin{equation}
\Gamma({\tilde \chi}^0_1 \rightarrow \nu \tau^{\pm}l^{\mp}) \sim
\frac{g'^2 h_{\tau}^2}{512 \pi^3}\Big(\frac{\vec \epsilon}{\mu}\Big)^2 
\Big(\frac{m_{{\tilde \chi}^0_1}}{m_{\tilde \tau}}\Big)^4  
g(\frac{m^2_{{\tilde \chi}^0_1}}{m^2_{\tilde \tau}})m_{{\tilde \chi}^0_1}.
\label{GamRPV3L}
\end{equation}
Here, 
\begin{equation}
g(y) = \frac{12}{y^2}(-5/2 +3/y + (-1 + 1/y) (-1 + 3/y) {\rm ln}(1-y) )
\label{defcorrstau}
\end{equation}
and we have normalized $g(y)$ conveniently such that $g(0) \rightarrow 1$, 
thus $g(y)$ varies between $[1,6]$. Note, that 
$\Gamma({\tilde \chi}^0_1 \rightarrow \nu \tau^{\pm}l^{\mp})$ goes to zero 
proportional to the fourth power of $m_{{\tilde \chi}^0_1}/m_{\tilde \tau}$. 
Since in GMSB the (right) scalar tau is never very much heavier than 
the neutralino, contrary to squarks and other scalar leptons, 
Eq.(\ref{GamRPV3L}) usually dominates over other Feynman graphs with 
scalar exchange.

Eq.(\ref{GamRPV3L}) can take a wide range of values. Just to give a 
flavour of its typical size, for $\frac{|\vec\epsilon|}{\mu} \sim 
3\times 10^{-4}$, $\tan\beta = 10$, and $m_{{\tilde \chi}^0_1} = 100$ GeV and 
$m_{\tilde \tau} = 120$ GeV one finds $\Gamma({\tilde \chi}^0_1 
\rightarrow \nu \tau^{\pm}l^{\mp}) \sim 3 \times 10^{-3}$ eV. We have 
checked numerically, see next section, that the decays described by 
eqs (\ref{GamRPV}) and (\ref{GamRPV3L}) are usually the most important 
RpV decay channels in GMSB.

The decay width of a neutralino NLSP to gravitino-photon is given by  
\cite{Giudice:1998bp}
\begin{eqnarray}\nonumber 
\Gamma({\tilde \chi}^0_1 \rightarrow {\tilde G}\gamma) &=& 
\frac{ \kappa_{\gamma}^2 m_{{\tilde \chi}^0_1}^5}{48 \pi m_{3/2}^2 M_{Pl}^2} \\
& \simeq & 
1.2 \times 10^{-6} \kappa_{\gamma}^2 \Big(\frac{m_{{\tilde \chi}^0_1}}
{\rm 100 \hskip2mm GeV}\Big)^5
\Big(\frac{\rm 100 \hskip2mm eV}{m_{3/2}}\Big)^{2} 
\hskip1mm {\rm eV}
\label{NLSPGrav}
\end{eqnarray}
Here, \footnote{In MSSM notation. In the notation of  \cite{hirsch:2000ef}: 
$N_{1j} \rightarrow N_{4j}$} 
$\kappa_{\gamma} = |\cos\theta_W N_{11} + \sin\theta_W N_{12}|$. Neutralinos 
can also decay into ${\tilde \chi}^0_1 \rightarrow {\tilde G}Z^0$ and 
${\tilde \chi}^0_1 \rightarrow {\tilde G}h^0$ and we include those channels 
in our numerical calculation. However, these final states are usually 
less important than ${\tilde \chi}^0_1 \rightarrow {\tilde G}\gamma$ 
and also do not give a promising signal. We will therefore not discuss 
them in further details. 

From eqs (\ref{GamRPV}), (\ref{GamRPV3L}) and (\ref{NLSPGrav}) we can very 
roughly estimate a branching ratio of 
$Br({\tilde \chi}^0_1 \rightarrow {\tilde G}\gamma) \sim 
10^{-(2-3)}$ for $m_{3/2}$ of ${\cal O}(100)$ eV. This is the main 
result of the current paper. We will back up this estimate with a 
numerically exact calculation in the next section.

We note for comparison that if R-parity is conserved, Eq.(\ref{NLSPGrav}) 
gives a typical decay length of $c\tau \sim 20 (\frac{m_{3/2}}{\rm keV})^2$ 
m, for a neutralino mass of $m_{{\tilde \chi}^0_1} \simeq 100$ GeV. According 
to \cite{Kawagoe:2003jv} the ATLAS detector at LHC should be able 
to measure even such a ``large'' decay length rather well given 
sufficient luminosity ($100 fb^{-1}$). 

Finally, we have checked that the gravitino itself lives long enough to 
be a dark matter candidate. Following \cite{Takayama:2000uz} the 
decay width of ${\tilde G}\rightarrow \nu \gamma$ can be calculated 
from the photino content of the neutrino as
\begin{equation}
\Gamma({\tilde G}\rightarrow \sum_i \nu_i \gamma) \simeq \frac{1}{32 \pi}
|U_{\gamma \nu}|^2 \frac{m_{3/2}^3}{M_{Pl}^2}.
\label{GravDec}
\end{equation}
Here, $|U_{\gamma \nu}|^2 = \sum_{i=1}^3 |\cos\theta_W N_{i1} + 
\sin\theta_W N_{i2}|^2$. The coupling matrices $N_{i1}$ and $N_{i2}$
can be calculated perturbatively \cite{hirsch:2000ef} and are 
approximately fixed from the neutrino masses. For $M_1= 100$ GeV and 
$m_{3/2} = 100$ eV, 
we find $|U_{\gamma \nu}|^2 \sim 3.5\times 10^{-14} \frac{m_{\nu}}
{0.05 \hskip1mm {\rm eV}}$. This corresponds to a half-life of the 
order of $10^{31}$ Hubble times.

\section{Numerical results}

We stress that none of the approximations discussed above were used in 
our numerical analysis. Numerical results presented below have been 
obtained as follows. We generated supersymmetric particle spectra 
using the package SPheno \cite{Porod:2003um}. GMSB is characterized 
by two mass scales, the scale of supersymmetry breaking $F$ and the 
messenger mass $M_M$. Since $F$ is related to the gravitino mass via 
\cite{Giudice:1998bp}
\begin{equation}
m_{3/2} = \frac{F}{k\sqrt{3}M_P}
\label{FtoGrav}
\end{equation}
we trade $F$ for the gravitino mass and vary $\Lambda^{GMSB}=F/M_M$ 
independently. For details see \cite{Giudice:1998bp}. For definiteness we 
take $k =1/20$. We have checked that our results depend only very weakly 
on the exact value of $k$ as long as $k < 1$. In addition, we have 
$\tan\beta$, the sign of $\mu$ and the number of messengers, $n_5$, as 
free parameters. We use only $n_5 = 1,2$ because for larger values 
of $n_5$ the neutralino is rarely the NLSP, since scalar masses scale as 
$m_{\tilde \tau}/m_{{\tilde \chi}^0_1} \sim 1/\sqrt{n_5}-b $.-A

We check the generated spectra for a number of phenomenological limits 
\cite{Eidelman:2004wy}: $m_{{\tilde \chi}^+_1} \ge 105$ GeV, 
$m_{\tilde \mu} \ge 95$ GeV and $m_{\tilde \tau} \ge 82$ GeV. Lower limits 
on the mass of the lightest neutralino in GMSB have been published by all 
LEP collaborations \cite{Abdallah:2003np}, bounds are between $92-100$ GeV 
depending on the details of the analysis. 
However, we have found that the most important constraint for us is the 
lower limit on the mass of the lightest Higgs.\cite{Barate:2003sz}, which 
essentially cuts out all points with $m_{{\tilde \chi}^0_1} \lsim 100$ GeV 
from our scans.

R-parity violation is then included into SPheno \cite{Porod:2003um} as 
described for neutrino masses in \cite{hirsch:2000ef} and for neutralino 
decays in \cite{Porod:2000hv}. Special care is taken to diagonalize the 
neutrino-neutralino mass matrix at 1-loop order to give neutrino 
masses and mixings compatible with the values indicated by atmospheric 
and solar neutrino experiments \cite{Maltoni:2004ei}. For each set of 
GMSB parameters, $m_{3/2}$, $\Lambda^{GMSB}$, $\tan\beta$, $sgn(\mu)$ 
and $n_5$, this results in a restricted range of $\Lambda_i$ and 
$\epsilon_i$, as discussed above, from which then the RpV neutralino 
decays are calculated. From the discussion in the previous section one 
expects that the errors in our calculated branching ratios scale 
proportional to the errors in the neutrino masses and we have 
checked numerically that this assertion is correct. Results discussed below 
use neutrino masses near the best fit points for solar and atmospheric 
physics \cite{Maltoni:2004ei}. Finally, for the 
calculation of the branching ratio into gravitino plus photon 
Eq. (\ref{NLSPGrav}) is used. 

Our numerical results show that the dominant final states are 
usually either $\tau^{\pm}l^{\mp}\nu$ or $W^{\pm}l^{\mp}$ (and $Z^0\nu$). 
Gauge boson final states become more important the larger 
$m_{\tilde \tau}-m_{{\tilde \chi}^0_1}$, as discussed above, and for 
larger $m_{{\tilde \chi}^0_1}$. We find that for $m_{{\tilde \chi}^0_1} 
\ge 150$ GeV the final state $h^0\nu$ can reach a branching ratio of the 
order of $5-15$ \%. All other final states usually have branching ratios 
which are smaller. Especially, we find that the final state $b{\bar b}\nu$ 
is less important than in an mSugra scenario \cite{Porod:2000hv}. This 
can be understood as being due to the smaller ratio of 
$m_{\tilde \tau_1}/m_{\tilde b_1}$ in GMSB compared to mSugra. 
With typical total widths in the range of roughly 
${\cal O}(10^{-4})-{\cal O}(10^{-2})$ eV, we expect that the neutralino 
decays with a displaced vertex. 

In Figure 1 we show the branching ratio $BR({\tilde \chi}^0_1 \rightarrow 
{\tilde G}\gamma)$ as a function of NLSP mass for $n_5=1$, two values of 
$\tan\beta$, both signs of $\mu$ and for a fixed value of $m_{3/2} = 0.1$ 
keV. The branching ratio rises strongly with increasing neutralino mass, 
as expected from eq. (\ref{NLSPGrav}). The plot also shows that the 
dependence on $\tan\beta$ is rather weak, changing $\tan\beta$ from 
10 to 35 induces a change in the branching ratio up to a factor of 
$\sim 2$. It is also obvious that the sign of $\mu$ is not decisive. 
Choosing $n_5=2$ reduces the branching ratio by typically a factor 
of $\sim 3$ compared to the results shown.

\begin{figure}[htbp]
\begin{center}
\vspace{5mm}
\includegraphics[width=80mm,height=8cm]{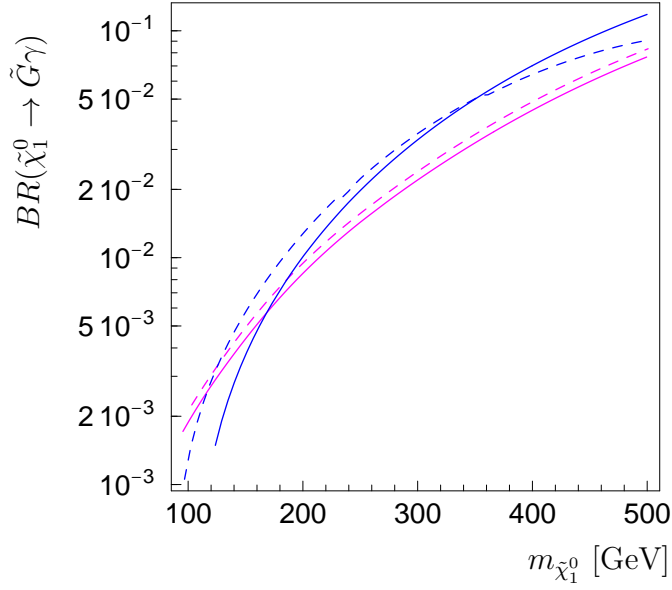}
\end{center}

\vskip-50mm\hskip25mm  
\begin{rotate}{90}
$BR({\tilde \chi}^0_1 \rightarrow {\tilde G}\gamma)$
\end{rotate}

\vskip40mm\hskip90mm $m_{{\tilde \chi}^0_1}$ [GeV]

\vspace{0mm}
\caption{$BR({\tilde \chi}^0_1 \rightarrow {\tilde G}\gamma)$ 
as function of the lightest neutralino mass, $m_{{\tilde \chi}^0_1}$ [GeV]. 
Full lines are for $\mu > 0$, dashed lines $\mu < 0$. Light (on colour 
printers magenta): $\tan\beta =10$, Dark (blue): $\tan\beta =35$.}
\label{fig:mNtrl}
\end{figure}

\begin{figure}[htbp]
\begin{center}
\vspace{5mm}
\includegraphics[width=80mm,height=80mm]{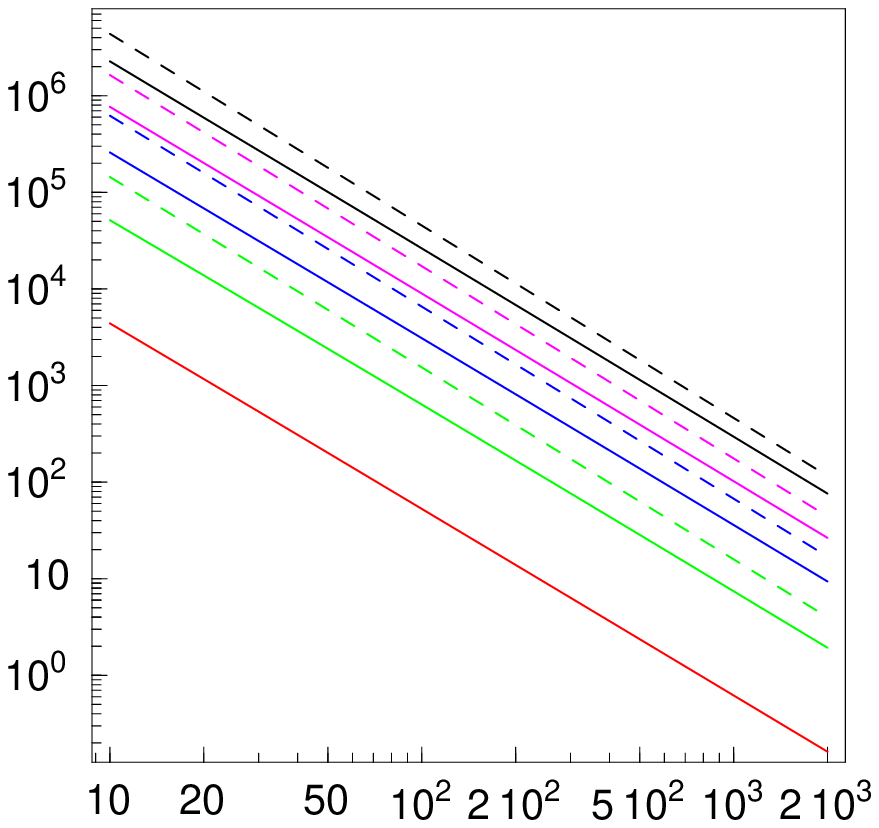}
\end{center}

\vskip-30mm\hskip25mm  
\begin{rotate}{90}
$BR({\tilde \chi}^0_1 \rightarrow {\tilde G}\gamma)/
BR({\tilde \chi}^0_1 \rightarrow \nu\gamma)$
\end{rotate}

\vskip25mm\hskip90mm $m_{3/2}$ [eV]

\vskip-35mm\hskip55mm {\tiny $m_{{\tilde \chi}^0_1} = 100$ GeV}
\vskip-35mm\hskip75mm {\tiny $m_{{\tilde \chi}^0_1} = 500$ GeV}

\vspace{60mm}
\caption{Ratio $BR({\tilde \chi}^0_1 \rightarrow {\tilde G}\gamma)/
BR({\tilde \chi}^0_1 \rightarrow \nu\gamma)$ as function of $m_{3/2}$ [eV]. 
Full lines are for $\tan\beta =10$, dashed lines for $\tan\beta =35$. The 
plots shows the case $n_5=1$, for $n_5=2$ the ratio is typically a 
factor (2-3) smaller. The different lines are for 
(from bottom to top) $m_{{\tilde \chi}^0_1} = 100-500$ GeV in steps 
of $100$ GeV.}
\label{fig:Background}
\end{figure}

The experimental signal for the final state ${\tilde G}\gamma$ is 
\eslash $\gamma$. In R-parity violating models the neutralino has 
another decay mode which gives the same experimental signal, namely 
${\tilde \chi}^0_1 \rightarrow \nu\gamma$. In BRpV this occurs at 1-loop 
order. To estimate this background we have done a 
calculation of  $BR({\tilde \chi}^0_1\rightarrow \nu\gamma)$. 
Fig. (\ref{fig:Background}) shows $BR({\tilde \chi}^0_1\rightarrow 
{\tilde G}\gamma)/BR({\tilde \chi}^0_1\rightarrow \nu\gamma)$ 
as a function of the gravitino mass, for two different choices of 
$\tan\beta$ and for $n_5=1$ for various different values of 
$m_{{\tilde \chi}^0_1}$. 
The ratio depends strongly on the neutralino and gravitino masses, for 
$m_{3/2} \le 500$ eV it is always larger than $1$. For neutralino 
masses greater than about $m_{{\tilde \chi}^0_1} \simeq 150$ GeV 
$\nu\gamma$ never 
seems to be a serious problem up to gravitino masses $m_{3/2} = 2$ keV.

\begin{figure}[htbp]
\begin{center}
\vspace{5mm}
\includegraphics[width=120mm,height=10cm]{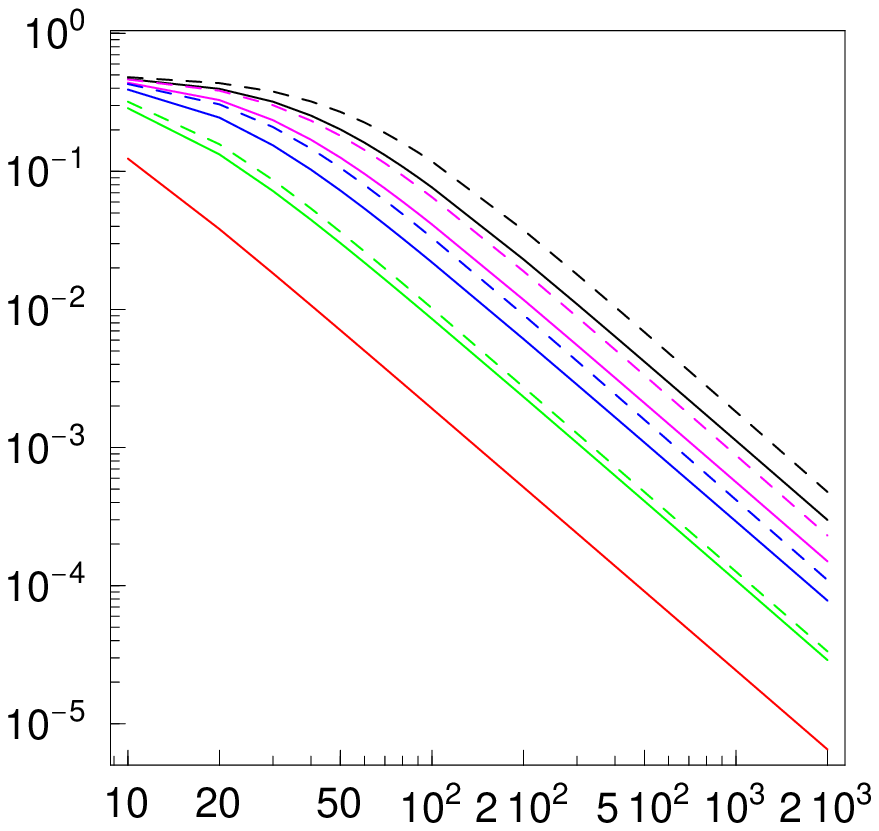}
\end{center}

\vskip-70mm\hskip5mm  
\begin{rotate}{90}
$BR({\tilde \chi}^0_1 \rightarrow {\tilde G}\gamma)$
\end{rotate}

\vskip65mm\hskip105mm $m_{3/2}$ [eV]

\vskip-55mm\hskip45mm {\small $m_{{\tilde \chi}^0_1} = 100$ GeV}
\vskip-35mm\hskip80mm {\small $m_{{\tilde \chi}^0_1} = 500$ GeV}

\vspace{80mm}
\caption{Ratio $BR({\tilde \chi}^0_1 \rightarrow {\tilde G}\gamma)$ 
as function of the gravitino mass, $m_{3/2}$ [eV]. Full lines are for 
$\tan\beta =10$, dashed lines $\tan\beta =35$. Lines from bottom to top 
are for different values of $m_{{\tilde \chi}^0_1}$, 
$m_{{\tilde \chi}^0_1} = 100-500$ GeV in steps of $100$ GeV. 
$BR({\tilde \chi}^0_1 \rightarrow {\tilde G}\gamma)$ is larger than 
$10^{-4}$ for values of $m_{3/2}$ between $0.5-2.0$ keV, 
depending on the neutralino mass.}
\label{fig:mGrav}
\end{figure}

Finally, fig. (\ref{fig:mGrav}) shows our main result. Here we plot 
$BR({\tilde \chi}^0_1 \rightarrow {\tilde G}\gamma)$ as a function 
of the gravitino mass for different values of $m_{{\tilde \chi}^0_1}$, 
two values of $\tan\beta$ and $n_5=1$. $n_5=2$ leads to branching ratios 
approximately a factor of up to $3$ smaller. Depending on the 
neutralino mass, $BR({\tilde \chi}^0_1 \rightarrow {\tilde G}\gamma)$ 
is larger than $10^{-4}$ for values of $m_{3/2} \simeq 0.5$ keV up to 
$m_{3/2} \simeq 2$ keV. At the LHC one expects to produce very roughly 
of the order of ${\cal O}(10^5)$ - ${\cal O}(10^7)$ events from 
supersymmetry, depending mainly on squark and gluino masses. Thus we 
think that sufficient statistics to measure branching ratios as small 
as $10^{-4}$ should be possible. We conclude therefore, that for 
cosmologically interesting ranges for the gravitino mass measurably 
large branching ratios $BR({\tilde \chi}^0_1 \rightarrow {\tilde G}\gamma)$ 
should exist.

\section{Conclusions}

We have discussed decay properties of the lightest neutralino in 
models with gauge mediated supersymmetry breaking in which the 
neutrino masses and mixings are explained by bilinear R-parity 
violation. Once the BRpV parameters are approximately fixed with 
information from the neutrino sector, the branching ratio into 
gravitino plus photon is fixed to be in the range $10^{-(2-3)}$ 
for a gravitino mass of the order of (few) $100$ eV. The branching ratio 
decreases with increasing gravitino mass. In the scenario discussed 
one can therefore test whether the gravitino gives a significant 
contribution to the dark matter of the universe by a ``simple'' 
counting experiment.

Can one do better? - Concerning the gravitino mass, the answer is yes. 
A measurement of the decay length of the neutralino 
would fix the total neutralino decay width, independent of our assumptions 
about neutrino masses. Knowing the width and the mass of the neutralino 
fixes the gravitino mass from the measurement of 
$Br({\tilde \chi}^0_1 \rightarrow {\tilde G}\gamma)$ in a much tighter range 
than what we have been able to do. However, one has to admit that even 
knowing the gravitino mass rather well, one can not calculate 
$\Omega_{3/2}$ reliably from $m_{3/2}$ without making specific assumptions 
on the thermal history of the universe. In this sense, gravitino DM can 
be ruled out by the measurement we have discussed, but never 
``experimentally confirmed''. 

\section*{Acknowledgments}
We thank Sergio Pastor and Giacomo Polesello for discussions. 
This work was supported by Spanish grant BFM2002-00345, by the
European Commission Human Potential Program RTN network
HPRN-CT-2000-00148 and by the European Science Foundation network
grant N.86.  M.H. and W.P. are supported by  MEC Ramon y Cajal contracts.
W.P.~is supported partly by the Swiss `Nationalfonds'. 
D.R is supported by UV postdoctoral grant associated to project
MEC-FPA2002-1149-E.


\begin{thebibliography}{10}

\bibitem{Giudice:1998bp}
For a review on gauge mediated supersymmetry breaking, see: 
G.~F.~Giudice and R.~Rattazzi,
Phys.\ Rept.\  {\bf 322} (1999) 419
[arXiv:hep-ph/9801271].

\bibitem{Pagels:1981ke}
H.~Pagels and J.~R.~Primack,
Phys.\ Rev.\ Lett.\  {\bf 48} (1982) 223.

\bibitem{Bertone:2004pz}
For a recent review on dark matter (with emphasis on neutralino 
detection): G.~Bertone, D.~Hooper and J.~Silk,
Phys.\ Rept.\  {\bf 405} (2005) 279
[arXiv:hep-ph/0404175].

\bibitem{hirsch:2000ef}
M.~Hirsch, M.~A. Diaz, W.~Porod, J.~C. Romao and J.~W.~F. Valle,
\newblock Phys. Rev. {\bf D62}, 113008 (2000), [hep-ph/0004115];
[Erratum-ibid.\ D {\bf 65} (2002) 119901];
M.~A. Diaz, M.~Hirsch, W.~Porod, J.~C. Romao and J.~W.~F. Valle,
{\bf D68}, 013009 (2003) \newblock [hep-ph/0302021] 

\bibitem{hempfling:1996wj}
R.~Hempfling,
\newblock Nucl. Phys. {\bf B478}, 3 (1996), [hep-ph/9511288];
D.~E. Kaplan and A.~E. Nelson,
\newblock JHEP {\bf 01}, 033 (2000), [hep-ph/9901254];
S.~Y. Choi, E.~J. Chun, S.~K. Kang and J.~S. Lee,
\newblock Phys. Rev. {\bf D60}, 075002 (1999), [hep-ph/9903465];
E.~J. Chun and S.~K. Kang,
\newblock Phys. Rev. {\bf D61}, 075012 (2000), [hep-ph/9909429];
E.~J. Chun,
\newblock Phys. Lett. {\bf B525}, 114 (2002), [hep-ph/0105157];
F.~Borzumati and J.~S. Lee,
\newblock Phys. Rev. {\bf D66}, 115012 (2002), [hep-ph/0207184];
A.~S. Joshipura, R.~D. Vaidya and S.~K. Vempati,
\newblock Nucl. Phys. {\bf B639}, 290 (2002), [hep-ph/0203182].

\bibitem{drees:1998id}
M.~Drees, S.~Pakvasa, X.~Tata and T.~ter Veldhuis,
\newblock Phys. Rev. {\bf D57}, 5335 (1998), [hep-ph/9712392];
A.~S. Joshipura and S.~K. Vempati,
\newblock Phys. Rev. {\bf D60}, 111303 (1999), [hep-ph/9903435];
A.~S. Joshipura, R.~D. Vaidya and S.~K. Vempati,
\newblock Phys. Rev. {\bf D65}, 053018 (2002), [hep-ph/0107204];
V.~D. Barger, T.~Han, S.~Hesselbach and D.~Marfatia,
\newblock Phys. Lett. {\bf B538}, 346 (2002), [hep-ph/0108261].

\bibitem{davidson:2000ne}
S.~Davidson and M.~Losada,
\newblock Phys. Rev. {\bf D65}, 075025 (2002), [hep-ph/0010325];
S.~Davidson and M.~Losada,
\newblock JHEP {\bf 05}, 021 (2000), [hep-ph/0005080];
E.~J. Chun, S.~K. Kang, C.~W. Kim and U.~W. Lee,
\newblock Nucl. Phys. {\bf B544}, 89 (1999), [hep-ph/9807327];

\bibitem{Maltoni:2004ei}For a recent review on neutrino oscillation 
data, see, for example: 
M.~Maltoni, T.~Schwetz, M.~A.~Tortola and J.~W.~F.~Valle,
New J.\ Phys.\  {\bf 6} (2004) 122
[arXiv:hep-ph/0405172].

\bibitem{Borgani:1996ag}
S.~Borgani, A.~Masiero and M.~Yamaguchi,
Phys.\ Lett.\ B {\bf 386} (1996) 189
[arXiv:hep-ph/9605222].

\bibitem{Takayama:2000uz}
F.~Takayama and M.~Yamaguchi,
Phys.\ Lett.\ B {\bf 485} (2000) 388
[arXiv:hep-ph/0005214].

\bibitem{Pierpaoli:1997im}
E.~Pierpaoli, S.~Borgani, A.~Masiero and M.~Yamaguchi,
Phys.\ Rev.\ D {\bf 57} (1998) 2089
[arXiv:astro-ph/9709047].

\bibitem{Eidelman:2004wy}
S.~Eidelman {\it et al.}  [Particle Data Group],
Phys.\ Lett.\ B {\bf 592} (2004) 1.

\bibitem{Primack:1997av}
J.~R.~Primack,
arXiv:astro-ph/9707285.

\bibitem{Moore:1999gc}
B.~Moore, T.~Quinn, F.~Governato, J.~Stadel and G.~Lake,
Mon.\ Not.\ Roy.\ Astron.\ Soc.\  {\bf 310} (1999) 1147
[arXiv:astro-ph/9903164].

\bibitem{Bode:2000gq}
P.~Bode, J.~P.~Ostriker and N.~Turok,
Astrophys.\ J.\  {\bf 556} (2001) 93
[arXiv:astro-ph/0010389].

\bibitem{Swaters:2002rx}
R.~A.~Swaters, B.~F.~Madore, F.~C.~V.~Bosch and M.~Balcells,
Astrophys.\ J.\  {\bf 583} (2003) 732
[arXiv:astro-ph/0210152].

\bibitem{Avila-Reese:2000hg}
V.~Avila-Reese, P.~Colin, O.~Valenzuela, E.~D'Onghia and C.~Firmani,
Astrophys.\ J.\  {\bf 559} (2001) 516
[arXiv:astro-ph/0010525].

\bibitem{Narayanan:2000tp}
V.~K.~Narayanan, D.~N.~Spergel, R.~Dave and C.~P.~Ma, 
Astrophys. J. {\bf 543} (2000) L103, 
[arXiv:astro-ph/0005095]

\bibitem{Viel:2005qj}
M.~Viel, J.~Lesgourgues, M.~G.~Haehnelt, S.~Matarrese and A.~Riotto,
arXiv:astro-ph/0501562.

%
\bibitem{Baltz:2001rq}
E.~A.~Baltz and H.~Murayama,
JHEP {\bf 0305} (2003) 067
[arXiv:astro-ph/0108172].

\bibitem{Fujii:2002fv}
M.~Fujii and T.~Yanagida,
Phys.\ Lett.\ B {\bf 549} (2002) 273
[arXiv:hep-ph/0208191].

%
\bibitem{Dimopoulos:1996gy}
S.~Dimopoulos, G.~F.~Giudice and A.~Pomarol,
Phys.\ Lett.\ B {\bf 389} (1996) 37
[arXiv:hep-ph/9607225].

\bibitem{Han:1997wn}
T.~Han and R.~Hempfling,
Phys.\ Lett.\ B {\bf 415} (1997) 161
[arXiv:hep-ph/9708264].

%
\bibitem{Carena:1997wy}
M.~Carena, S.~Pokorski and C.~E.~M.~Wagner,
Phys.\ Lett.\ B {\bf 430} (1998) 281
[arXiv:hep-ph/9801251].

\bibitem{Porod:2000hv}
W.~Porod, M.~Hirsch, J.~Romao and J.~W.~F.~Valle,
Phys.\ Rev.\ D {\bf 63} (2001) 115004
[arXiv:hep-ph/0011248].

\bibitem{Hirsch:2002ys}
M.~Hirsch, W.~Porod, J.~C.~Romao and J.~W.~F.~Valle,
Phys.\ Rev.\ D {\bf 66} (2002) 095006
[arXiv:hep-ph/0207334].

\bibitem{Kawagoe:2003jv}
K.~Kawagoe, T.~Kobayashi, M.~M.~Nojiri and A.~Ochi,
Phys.\ Rev.\ D {\bf 69} (2004) 035003
[arXiv:hep-ph/0309031].


\bibitem{Porod:2003um}
W.~Porod,
Comput.\ Phys.\ Commun.\  {\bf 153} (2003) 275
[arXiv:hep-ph/0301101].

\bibitem{Abdallah:2003np}
J.~Abdallah {\it et al.}  [DELPHI Collaboration],
Eur.\ Phys.\ J.\ C {\bf 38} (2005) 395
[arXiv:hep-ex/0406019];
A.~Heister {\it et al.}  [ALEPH Collaboration],
Eur.\ Phys.\ J.\ C {\bf 25} (2002) 339
[arXiv:hep-ex/0203024];
G.~Abbiendi {\it et al.}  [OPAL Collaboration],
Phys.\ Lett.\ B {\bf 602} (2004) 167
[arXiv:hep-ex/0412011];
P.~Achard {\it et al.}  [L3 Collaboration],
Phys.\ Lett.\ B {\bf 587} (2004) 16
[arXiv:hep-ex/0402002].

\bibitem{Barate:2003sz}
R.~Barate {\it et al.}  [ALEPH Collaboration],
Phys.\ Lett.\ B {\bf 565} (2003) 61
[arXiv:hep-ex/0306033].



\end{thebibliography}
\end{document}